\documentclass[12pt,a4paper]{article}
\usepackage{amsmath}
\usepackage{bm}
\usepackage{amsfonts}
\usepackage{graphicx}
\usepackage[normal]{caption2}
\usepackage{subfigure}
\usepackage{rotating}
\usepackage{citesort}
\usepackage{lscape}
\usepackage{section}
\begin{document}
\renewcommand\arraystretch{1.1}
\setlength{\abovecaptionskip}{0.1cm}
\setlength{\belowcaptionskip}{0.5cm}
\title {From fusion to total disassembly: global stopping in heavy-ion
collisions}
\author {Jatinder K. Dhawan, Narinder Dhiman, Aman D. Sood, and Rajeev K.
Puri\\
\it Department of Physics, Panjab University, Chandigarh -160 014,
India.\\}
 \maketitle
\begin{abstract}
Using the quantum molecular dynamics model, we aim to investigate
the emission of light complex particles, and degree of stopping
reached in heavy-ion collisions. We took incident energies between
50 and 1000 MeV/nucleon. In addition, central and peripheral
collisions and different masses are also considered. We observe
that the light complex particles act in almost similar manner as
anisotropic ratio. In other words, multiplicity of light complex
particles is an indicator of global stopping in heavy-ion
collisions. We see that maximum light complex particles and
stopping is obtained for heavier masses in central collisions.
\end{abstract}
 Electronic address:~rkpuri@pu.ac.in
\newpage
The study of nuclear reactions from low to relativistic energies
provides variety of phenomena. At low energies, Pauli principle
blocks any significant scattering of the nucleons. Therefore,
attractive mean field dominates the physics in this energy regime
\cite{lowener}. At intermediate energies, however, a mixture of
attractive mean field and repulsive nucleon-nucleon scattering
exists \cite{inter,aich}. Both these regimes together, lead the
matter from a fused state to total disassembly. One is also
interested to understand the mechanism behind this. Further, the
origin of small pieces (fragments) is also of great interest. One
is trying to correlate this origin with global stopping
\cite{goss}. The degree of stopping however, may vary drastically
with incident energies, mass of colliding nuclei and colliding
geometry. The degree of global stopping has also been linked with
the thermalization (equilibrium) in heavy-ion collisions.
\par
Theoretically, these happenings are followed by a variety of
models. Some models assume a {\it priori} equilibrium (at least at
local level) whereas others hunt for the degree of thermalization
in a reaction. Several models which depend on the assumption of
equilibrium, have been applied successfully to study the physics
at low and intermediate energies \cite{bond,moret}. At the same
time, the light and medium mass fragments (produced and emitted in
reactions), have also been used to get information about the
thermalization and stopping in heavy-ion collisions
\cite{goss,luka}. The origin of light and medium mass fragments is
still under debate \cite{aich,goss}.

\par
We here plan to investigate the degree of stopping reached in
these reactions from fusion to total disassembly. We shall also
attempt to correlate the emission of light mass fragments and
degree of stopping reached in a reaction. A complete knowledge
about the degree of stopping is very important since it can be
connected to the properties of the system equation of states and
in medium properties of the nucleon-nucleon cross section
\cite{wolter}. Our study is based on the analysis of reactions
between 50 and 1000 MeV/nucleon. Our aim is to look for the
randomization of one-body momentum space or memory loss of the
incoming momentum. This is also termed as global stopping in the
literature. Sometimes, this randomization is also related to the
dynamical thermalization of the nuclear matter in heavy-ion
collisions. For this analysis, we employ the quantum molecular
dynamics (QMD) model. The reader is referred to Ref.(3) for
details.
\par
For the present analysis, thousands of events were simulated for
the reactions of $^{20}$Ne+$^{20}$Ne, $^{40}$Ca+$^{40}$Ca,
$^{58}$Ni+$^{58}$Ni, $^{93}$Nb+$^{93}$Nb, $^{139}$La+$^{139}$La
and $^{197}$Au+$^{197}$Au between 50 and 1000 MeV/nucleon using
hard equation of state along with free energy dependent
nucleon-nucleon cross-section. The geometrical choice is varied
between the most central to peripheral ones. This large choice
will give us possibility to look for the matter from the fusion to
complete disassembly. The choice of symmetric reactions is to
avoid any influence of the asymmetry of colliding nuclei
\cite{jsingh}. As reported in Ref. \cite{jsingh}, the asymmetric
nuclei have quite different picture compared to symmetric nuclei.
One should keep in the mind that at low energies, one has
pre-equilibrium emission of particles from excited compound
nucleus.
\par
The colliding nuclei not only compress each other, they also heat
the matter \cite{goss,khoa}. In addition, the destruction of
initial correlations, makes the matter homogenous and one can have
global stopping. More the initial memory of nucleons is erased,
better it is stopped and better one has average mixing of
projectile and target momentum. We shall here consider few
different quantities capable of estimating the degree of global
stopping. The first quantity is the anisotropy ratio $<R_{a}>$
which is defined as \cite{khoa,puri}:
\begin{equation}
{\langle R_{a} \rangle = \frac{\sqrt{\langle p_{x}^{2}
\rangle}+\sqrt{\langle p_{y}^{2}\rangle}}{2\sqrt{\langle p_{z}^{2}
\rangle}}}.
\end{equation}

This anisotropy ratio $\langle R_{a} \rangle$ is an indicator of
the global stopping and randomization of momentum of target and
projectile of the system. The global word is due to the fact that
it does not depend on the local position. Naturally for a complete
mixing, $\langle R_{a} \rangle$ ratio should be close to unity.
Alternatively, in the literature, $\langle E_{rat} \rangle$
$\left(=\frac{\sum p_{\bot}^{2}/2mn}{\sum p_{z}^{2}/2mn}\right)$
has also been proposed for the mixing. There maximum value of
$\langle E_{rat} \rangle$ was supposed to be equal to 2
\cite{goss}.
\par

The second possibility for the degree of stopping of nuclear
matter is to look for the rapidity distribution which is defined
as
\begin{equation}
Y(i)= \frac{1}{2}\ln\frac{{\bf{E}}(i)+{\bf{p}}_{z}(i)}
{{\bf{E}}(i)-{\bf{p}}_{z}(i)},
\end{equation}
where ${{\bf E}(i)} $ and $ {\bf p}_{z}(i)$, are, respectively,
the total energy and longitudinal momentum of $ith$ particle.
Naturally, for a complete stopping, one should expect a single
Guassian shape of the rapidity. Very often, the nature of emitting
source is defined by analyzing the rapidity distribution.
\begin{figure}
\centering \vskip -2cm
\includegraphics[width=12cm]{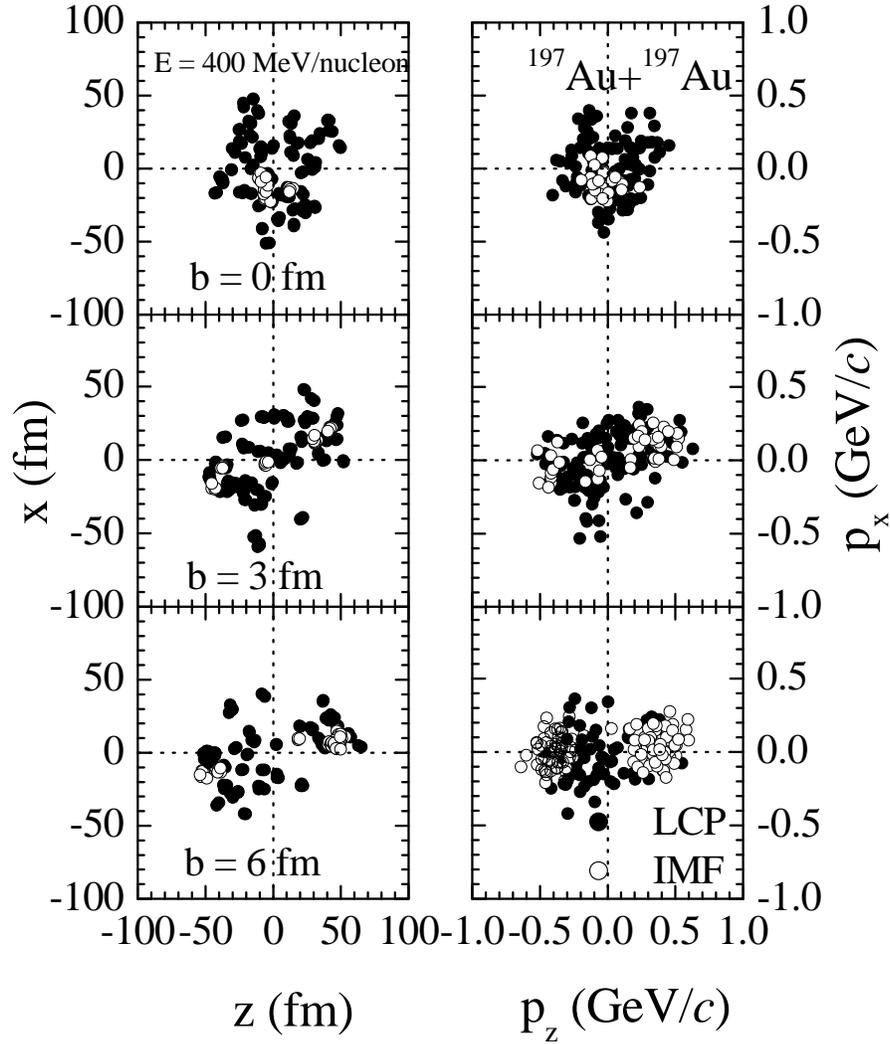}
\vskip - 1.0cm
 \caption{The snap-shots of a single event in the
phase-space. The top, middle, and bottom panels are, respectively,
for impact parameters b = 0, 3, and 6 fm. The filled circles
represent the LCP's and the open circles represent IMF's.}
\end{figure}
\begin{figure}
\centering \vskip -2cm
\includegraphics[width=12cm]{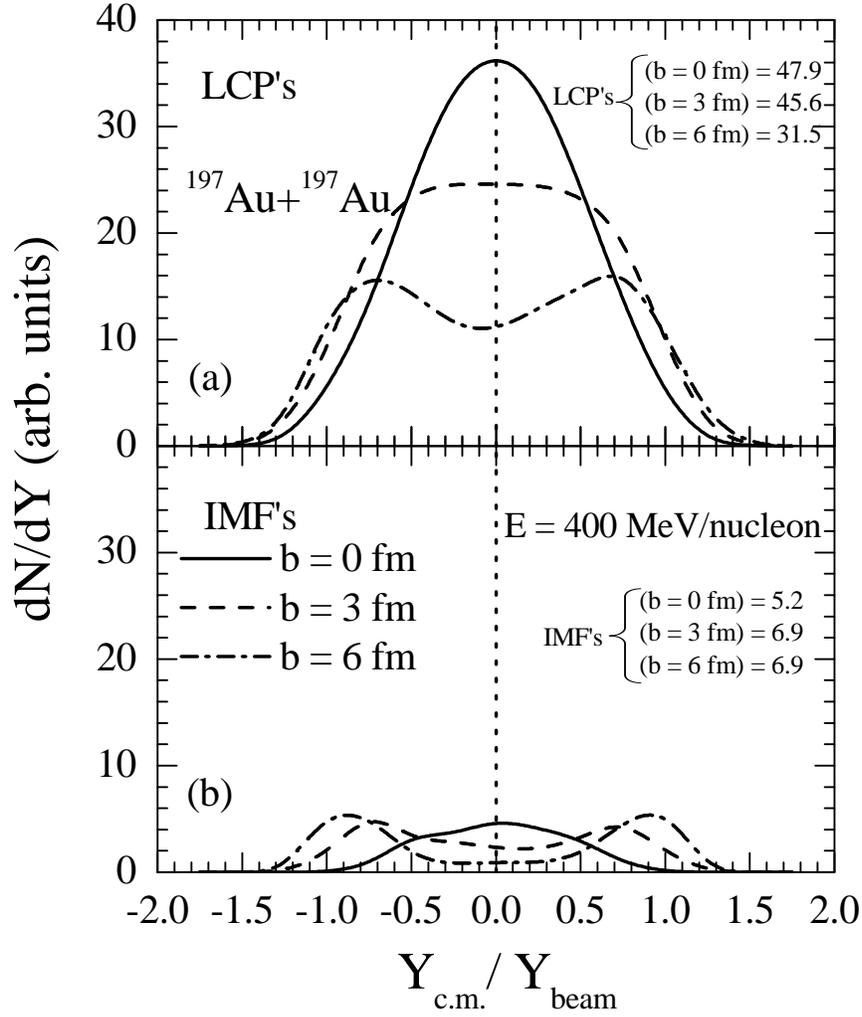}
\vskip - 1.0cm
 \caption{The rapidity distribution (of LCP's and
IMF's) $\frac{dN}{dY}$ as a function of reduced rapidity. The
reaction is of $^{197}$Au+$^{197}$Au at an incident energy of 400
MeV/nucleon.}
\end{figure}
\begin{figure}
\centering \vskip -2cm
\includegraphics[width=12cm]{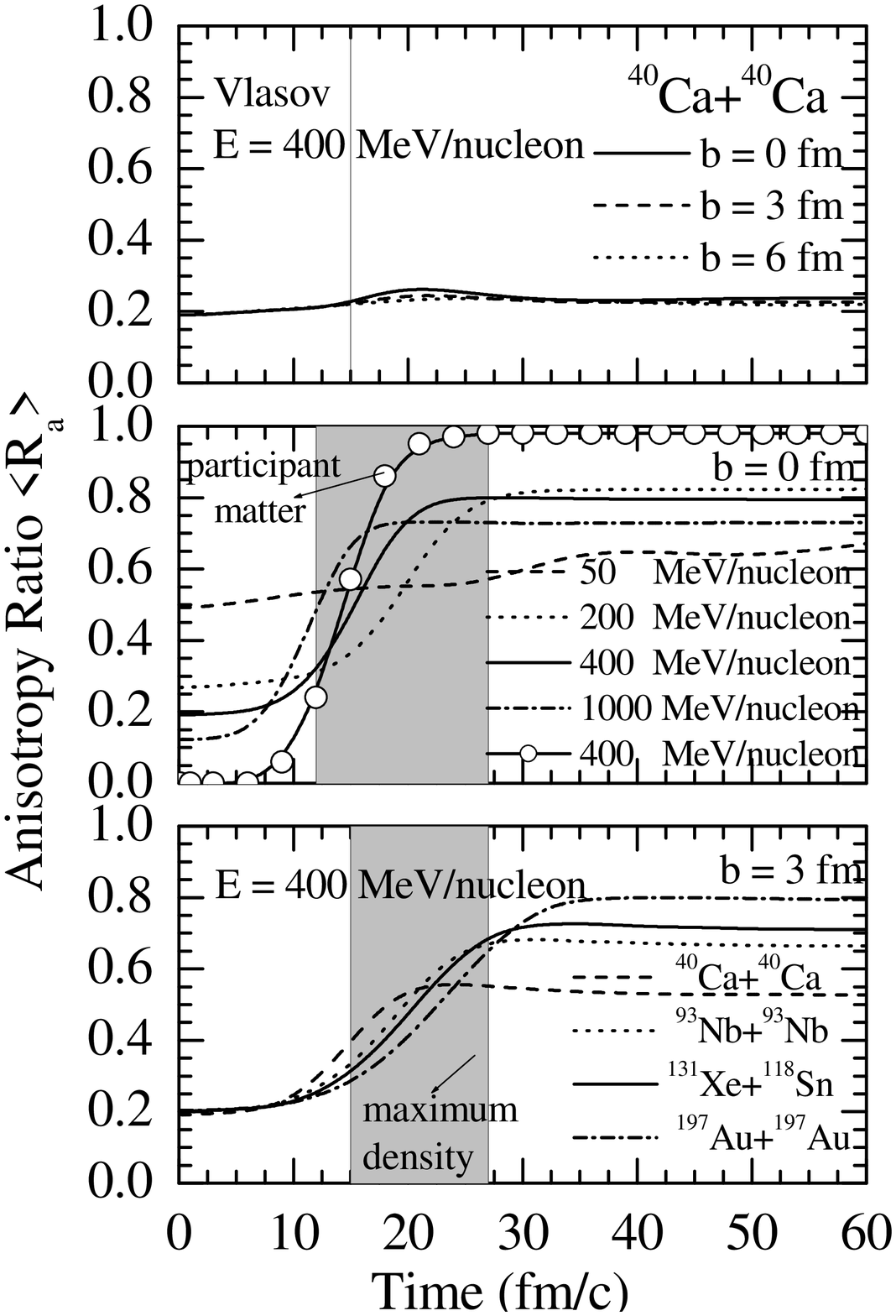}
\vskip - 0.3cm
 \caption{The time evolution of anisotropy ratio
$\langle R_{a} \rangle$ is displayed. The upper panel is for
$^{40}$Ca+$^{40}$Ca at an incident energy of 400 MeV/nucleon in
the Vlasov mode. The middle panel is for $^{40}$Ca+$^{40}$Ca at
different incident energies. The lower panel is for
$^{40}$Ca+$^{40}$Ca, $^{93}$Nb+$^{93}$Nb, $^{131}$Xe+$^{118}$Sn
and $^{197}$Au+$^{197}$Au, respectively.}
\end{figure}
\par
Apart from the global stopping and randomization of phase-space,
one may also define the local average mixing of target and
projectile. Here, one studies, the relative momentum of two
colliding Fermi spheres which indicates the deviation from a
single Fermi sphere and by that from mixing. The concept of local
stopping is used by the hydrodynamical models to simulate the
heavy-ion reactions. All these definitions will give us a glimpse
of the degree of stopping or mixing (and sometimes of equilibrium)
reached in a reaction. We shall also work out its relation with
the production of light complex particles (LCP's) defined as
fragments with $(2\leq A \leq 4)$. Here the light complex
particles are detected with two different methods: In the first
approach, one binds nucleons if they are within a distance of 4
fm. This method is labelled as minimum spanning tree (MST) method.
This method, by virtue, of it's definition may miss the light
complex particles emitted at a later stage of the reaction either
from quasi target/projectile or fusion residue. To cope with this,
we also employ the sophisticated simulated annealing
clusterization algorithm (SACA) as a second approach, which is
based on the idea of obtaining that fragment's structure which
maximizes the binding energy of the system \cite{saca}. This
algorithm is based on the Metropolis algorithm and has been
reported to explain the data throughout the energy range
\cite{saca}.
\par
In Fig. 1, we display the final phase-space of a single event of
$^{197}$Au+$^{197}$Au at 400 MeV/nucleon. The different panel are
at b = 0, 3 and 6 fm, respectively. Here only LCP's and IMF's
$(5\leq A \leq 65)$ are displayed. We note that the central
collisions lead to a complete spherical distribution. This degree
of stopping decreases with the increase in impact parameter.
Further the light complex particles mostly originate from the
mid-rapidity region. In other words, LCP's can also act as a
barometer for studying the stopping in heavy-ion collisions. The
IMF's, however, are pointing towards either target or projectile
regions. Therefore, these are originating from the surface of
colliding nuclei. In other words, these can be viewed as remnant
of the spectator matter. It has been discussed by many authors
that the intermediate mass fragments carry the initial memory of
nuclei and correlations. Naturally, one can not expect them to be
emitted from mid-rapidity region. Since degree of spectator matter
increase with impact parameter, so is the formation of either
IMF's or heavier fragments. We have also studied a large number of
different individual events. The above analysis is quite similar
in all these events.

\par
To further quantify this observation, we display in Fig. 2, the
rapidity distribution dN/dY for LCP's and IMF's. We see that the
LCP's emitted in central collisions are originating from a
complete stopped source. The central collisions are better
randomized compared to peripheral ones. The conclusions here match
with the Fig.1 and also in agreement with earlier reports
\cite{goss,moret,luka}. It is worth mentioning that the Gaussian
shape depends on the size of colliding nuclei. For lighter
colliding nuclei, it is more of a flat distribution whereas for
heavier nuclei, it is more of a Gaussian type \cite{jsingh}.

\begin{figure}
\centering \vskip -2cm
\includegraphics[width=12cm]{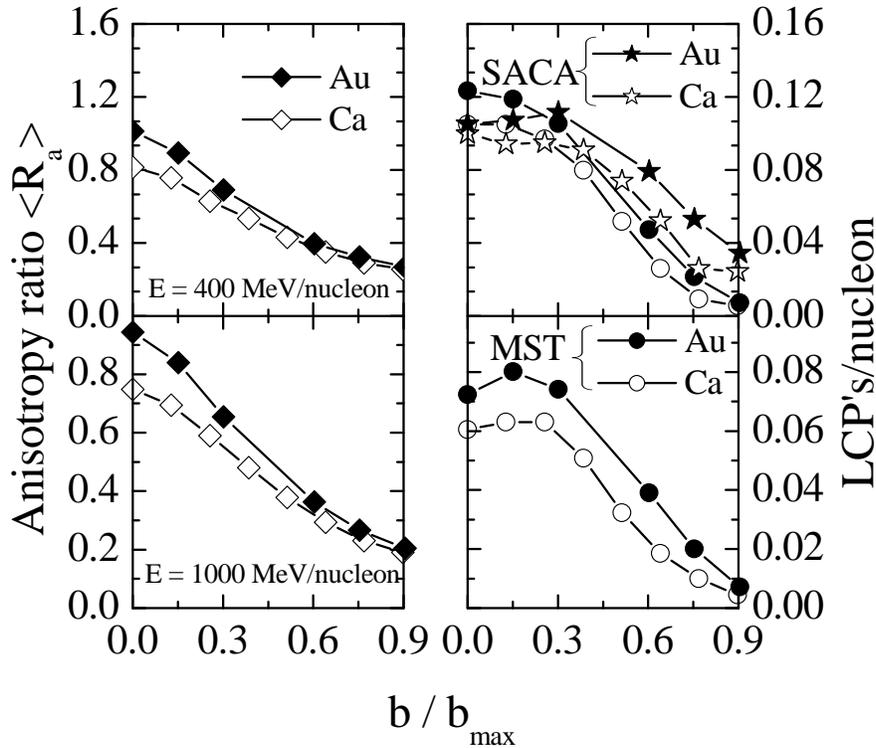}
\vskip -3.7 cm
 \caption{The anisotropic ratio $\langle R_{a} \rangle$
and LCP's/nucleon as a function of normalized impact parameter.
The reactions considered are of $^{40}$Ca+$^{40}$Ca and
$^{197}$Au+$^{197}$Au. The upper panel is at E = 400 MeV/nucleon
whereas lower panel is at E = 1000 MeV/nucleon. Here LCP's are
obtained within MST and SACA approaches.}
\end{figure}
\par
In Fig. 3 we display the global stopping and randomization in
terms of anisotropic ratio $\langle R_{a} \rangle$ in the
mass-energy-impact parameter plane. If we don't allow the
nucleon-nucleon collisions to happen (i.e. in a Vlasov mode),
matter does not randomize globally at all. It is argued by many
authors that the role of mean field is independent of mass of the
colliding system. Here we show that it is also independent of the
impact parameter. One notices that even at small incident energies
(e.g. 50 MeV/nucleon), 60$\%$ randomization can be achieved.
However, at very high incident energy (e.g. 1 GeV/nucleon), system
does not advance towards global randomization. Instead, a decrease
in the degree of stopping points toward transparency in nuclear
matter with increase in incident energy. This is also in agreement
with Ref. \cite{goss}.
\par

Looking at the bottom panel, one also notices a linear trend for
the mass dependence of anisotropic ratio $\langle R_{a} \rangle$.
At a fixed impact parameter, heavier colliding nuclei are better
randomized compared to lighter colliding nuclei. We will come back
to this point later on. This conclusion is independent of the mass
of colliding nuclei, impact parameter as well as incident energy.
From the Figure, one also notices that anisotropic ratio $\langle
R_{a} \rangle$ changes drastically during high density phase
($\rho>\rho_{0}$). Once high density phase is over, no more
changes occur in the thermalization. In other words, the
nucleon-nucleon collisions happening after the high density phase
do not produce substantial changes. It is worth mentioning that
collective transverse flow also saturates once high density phase
is over. Further, as shown in the Figure, participant matter
(nucleons suffering at least one collision) also follows the same
trend. It means the nucleons going under nucleon-nucleon
collisions after high density phase do not play a role. We shall
now attempt to correlate the emission of LCP's and anisotropic
ratio $\langle R_{a} \rangle$. As noted, both these are the
indicator of global stopping and randomization of momentum-space
in heavy-ion collisions.
\par
In Fig. 4, we display $\langle R_{a} \rangle$ ratio and
LCP's/nucleon (obtained in MST and SACA approaches) as a function
of the colliding geometry. Both $\langle R_{a} \rangle$ ratio and
LCP's/nucleon decrease (almost in same fashion) with increase in
impact parameter. Further the mass dependence has a little role to
play. A change of mass from 40 units to 400 units does not yield
significant changes in both the case. The same trend at both
incident energies points toward linear relation between both these
quantities. As stated earlier, the emission of LCP's originate
from a complete randomized source. It is also interesting  to note
that the trend of LCP's/nucleon is similar in both MST and SACA
methods pointing towards complete emission of LCP's in these
reactions. The decrease in LCP's complements corresponding
increase in IMF's and heavy fragments. These fragments are the
remanent of spectator matter which is non-stopped. Therefore, a
decrease in LCP's multiplicity directly measures a decrease in the
degree of randomization and hence global stopping. Therefore,
there is one to one correspondence between both these quantities.
\begin{figure}
\centering \vskip -2cm
\includegraphics[width=12cm]{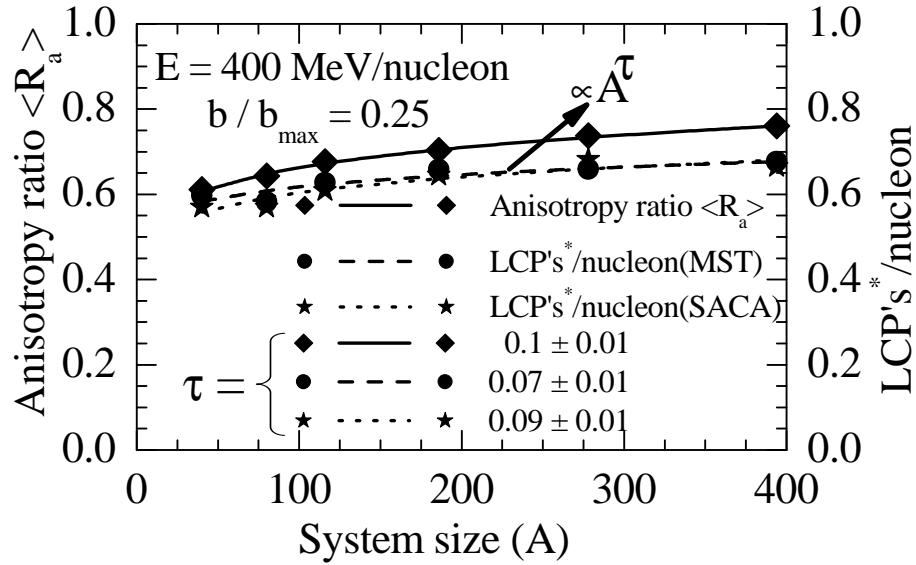}
\vskip -5.2 cm \caption{The anisotropy ratio $\langle R_{a}
\rangle$ and LCP's$^{*}$/nucleon as a function of composite mass
of colliding nuclei. Here we consider $^{20}$Ne+$^{20}$Ne,
$^{40}$Ca+$^{40}$Ca, $^{58}$Ni+$^{58}$Ni, $^{93}$Nb+$^{93}$Nb,
$^{139}$La+$^{139}$La and $^{197}$Au+$^{197}$Au reactions.}
\end{figure}
\par
In Fig. 5 we display the mass dependence of $\langle R_{a}
\rangle$ ratio and LCP's$^{*}$/nucleon (obtained in both MST and
SACA approaches) at 400 MeV/nucleon. Here, LCP's$^{*}$/nucleon are
scaled values of LCP's/nucleon by a factor of 6 ( i.e.
LCP's/nucleon*6). We see that $\langle R_{a} \rangle$ ratio has a
little mass dependence. Also, LCP's$^{*}$/nucleon in both
approaches has similar mass dependence. The mass dependence in
both the cases can be parameterized by a power law
$\propto$$A^{\tau}$ with $\tau$ close to 0.1, 0.07 and 0.09
respectively. A very similar impact parameter and mass dependence
of anisotropy ratio and LCP's$^{*}$/nucleon clearly demonstrate
that the LCP's production is closely related to the global
randomization in heavy-ion collisions.

As seen in Fig. 1, the LCP's are emitted from mid-rapidity region
where initial correlation and memory of nucleons is completely
destroyed. The $\langle R_{a} \rangle$ ratio is also a direct
indicator of breaking the initial correlations and erasing the
memory of nucleons. We see that the geometrical and mass behavior
is quite similar in both cases. Therefore, both, in principle are,
general quantities to study the global stopping. In addition, one
also notices that best randomization and stopping is obtained in
central collisions for heavier masses. If one takes impact
parameter averaged randomization, on notices that the this is
close to 0.6. One can say that no global stopping is reached in
inclusive heavy-ion collisions.
\par
In summary, using quantum molecular dynamics model, we investigate
the emission of light complex particles, and the degree of
stopping reached in heavy-ion collisions. We observed that the
light complex particles act in almost similar manner as
anisotropic ratio. In other words, multiplicity of light complex
particles production is an indicator of randomization in heavy-ion
collisions. We see that maximum light complex particles and
randomization is obtained for heavier mass at central collisions.
The smaller mass and larger impact parameters lead to less
equilibrated matter.

The work is supported by Department of Atomic Energy, Government
of India, India.



\begin{thebibliography}{999}

\bibitem{lowener} L. C. Vaz {\it et al.,} Phys. Rep. {\bf 69,} 373 (1981); R. K. Puri, Ph. D. Thesis, Panjab
University, Chandigarh (1990).
\bibitem{inter} H. St\"ocker, and W. Greiner, Phys. Rep. {\bf 137,}
277 (1986).
\bibitem{aich} J. Aichelin, Phys. Rep. {\bf 202,} 233
(1991); C. Hartnack {\it et al.}, Eur. Phys. J A {\bf 1,} 151
(1998).
\bibitem{goss} P. B. Gossiaux and J. Aichelin, Phys. Rev. C {\bf
56,} 2109 (1997).
\bibitem{bond} J. P. Bondorf {\it et al.}, Nucl. Phys. {\bf A443,} 321 (1985);
D. H. E. Gross, Rep. Prog. Phys. {\bf 53,} 605 (1990).
\bibitem{moret} L. G. Moretto and G. J. Wozniak, Annu. Rev. Nucl. Part. Sci. {\bf 43,} 379
(1993).
\bibitem{luka} J. Lukasik {\it et al.}, Phys. Rev. C {\bf 55,}
1906 (1997); {\it ibid}, Phys. Rev. C {\bf 61,} 014606 (1997);
{\it ibid}, Phys. Rev. C {\bf 66,} 064606 (2002); T. Lefort {\it
et al.}, Nucl. Phys. {\bf A602,} 397 (2000); F. Rami {\it et al.},
Phys. Rev. Lett. {\bf 84,} 1120 (2000); W. Reisdorf {\it et al.},
Phys. Rev. Lett. {\bf 92,} 232301 (2004).
\bibitem{wolter} B. Hong {\it et al.}, Phys. Rev. C {\bf 71,} 034902
(2005); T. Gaitanos, C. Fuchs and H. H. Wolter, Phys. Lett. B {\bf
609,} 241 (2005); P. Danielewicz, Acta Phys. Polon. B {\bf 33,} 45
(2002).

\bibitem{jsingh} J. Singh, S. Kumar and R. K. Puri, Phys. Rev. C {\bf 62,} 044617
(2000); {\it ibid}, Phys. Rev. C {\bf 65,} 024602 (2002).
\bibitem{khoa} D. T. Khoa {\it et al.}, Nucl. Phys. {\bf A542,}
671 (1992); R. K. Puri {\it et al.}, Nucl. Phys. {\bf A575,} 733
(1994); D. Hahn and H. St\"ocker, Nucl. Phys. {\bf A476,} 718
(1988).
\bibitem{puri} R. K. Puri {\it et al.}, J. Phys. G: Nucl. Part.
Phys. {\bf 20}, 1817 (1994).
\bibitem{saca} R. K. Puri {\it et al.}, Phys. Rev. C {\bf 54}, R28
(1996).
\end{thebibliography}
\end{document}